\documentclass[prc,preprint]{revtex4}
\usepackage{graphicx,float}
\usepackage{amssymb,amsmath}

\begin{document}
\title{Saturation properties of nuclear matter in the presence of strong magnetic field}

\author{ Z. Rezaei$^1$ and G. H. Bordbar$^{1,2}$\footnote{E-mail: ghbordbar@shirazu.ac.ir}} 

\affiliation{$^1$Department of Physics,
Shiraz University, Shiraz 71454, Iran \\ and \\ $^{2}$Center for Excellence in Astronomy and Astrophysics (CEAA-RIAAM)-Maragha,
P.O. Box 55134-441, Maragha 55177-36698, Iran}

\begin{abstract}
Different saturation properties of cold symmetric nuclear matter
in the strong magnetic field have been considered. We have seen that for magnetic fields about $B>  3 \times 10 ^ {17}\ G$,
{for
both cases with and without nucleon anomalous magnetic moments}, the saturation density and saturation energy
grow by increasing the magnetic field. It is indicated that the magnetic susceptibility of symmetric nuclear
matter becomes negative showing the diamagnetic response especially at $B<  3 \times 10 ^ {17}\ G$.
We have found that for the nuclear matter, the magnitude of orbital magnetization
reaches the higher values comparing to the spin magnetization. Our results for the incompressibility show that at high enough magnetic fields, i.e. $B>  3 \times 10 ^ {17}\ G$, {the softening of equation of state caused by Landau quantization is overwhelmed by stiffening due to the magnetization of nuclear matter.} We have shown that the effects of strong magnetic field on nuclear matter may affect the constraints on the equation of state of symmetric nuclear matter obtained applying
the experimental observable.
%
\end{abstract}
\maketitle
\section{Introduction}
\label{intro}
{Compact stars containing high-density nuclear matter with strong magnetic fields
are of the considerable interests in astrophysics. The magnetic field
of neutron stars has been estimated to be in the range $10^{11} - 10^{13}\ G$ in radio pulsars
\cite{Chashkina} and $10^{15} - 10^{16}\ G$ in magnetars \cite{Dvornikov}.
By comparing with the observational data, Yuan et al. obtained a magnetic field
strength of order $10^{19}\ G$ for neutron stars \cite{Yuan98}.
The magnetic field of the neutron stars is believed to have a dipole configuration \cite{Horowitz}.
The magnetic field could also originate in the creation of neutron stars due to the conservation of magnetic flux.
In addition, rapid and differential rotation of the collapsing core could amplify the magnetic field \cite{Endeve,Kiuchi}.
It has been shown that the highly twisted magnetar magnetospheres have influences on
the dipolar magnetic fields \cite{Beloborodov,Elfritz}.
Dynamo process \cite{Duncan}, magneto-rotational instabilities
\cite{Guilet}, Tayler-Spruit dynamo \cite{Tayler,Spruit}, and post-infall convection \cite{Obergaulinger} could also amplify the
magnetic field.} Strongly interacting nuclear matter under extreme conditions
can also be studied using the experiments on the heavy
ion collisions \cite{Hermann}. Heavy ion collisions with the energy up
to $10\ GeV$ provide the compressed nuclear matter to high
density \cite{Danielewicz}. In addition, the strong magnetic field is also
created in heavy ion collisions \cite{Kharzeev,Skokov,Bzdak2}. The energetic collisions
of heavy ions provide significant constraints on the
equation of state (EOS) of nuclear matter \cite{Danielewicz}.

These very intense magnetic fields change the properties
of symmetric nuclear matter. Some studies have investigated
the properties of neutron star matter \cite{Brod0,zhang,Yue},
quark matter \cite{Menezes,Chu,Bandyopadhyay}, and
nuclear matter \cite{Chakra7,Diener,Aguirre,Lopes} in the strong magnetic field.
In a relativistic Hartree theory, cold symmetric nuclear matter
and nuclear matter in beta equilibrium were probed in the strong magnetic
field without considering the effects of anomalous magnetic moments \cite{Chakra7}.
Their results showed that the application of magnetic field leads to the additional binding for
the system with a softer equation of state. They found that the saturation density of
nuclear matter increases by increasing
the magnetic field.
Within the context of effective baryon-meson exchange models with magnetic field
coupled to the charge and the dipole moment of the baryons, the symmetric nuclear matter
was investigated by Diener and Scholtz \cite{Diener}.
They found that by increasing the magnetic field, the saturation density of nuclear matter increases, while the system becomes less bound.
In addition, they have also shown that as the magnetic field and consequently the saturation density increase, the system becomes more incompressible.
Using non-relativistic
Skyrme potential model within a Hartree-Fock approach, the equation of state of nuclear matter in the presence of a magnetic field has been considered \cite{Aguirre}. By studying the compressibility and magnetic susceptibility, they have investigated the effect of coexistence of phases on the equilibrium configuration.

In this paper, we intend to investigate the saturation properties of symmetric nuclear
matter at zero temperature in the presence of strong magnetic field
using the lowest order constrained variational (LOCV) technique \cite{Bordbar23,Bordbar57,Bordbar76,Bordbar80,Bordbar83,Bordbar718} employing $AV_{18}$ inter-nucleon potential.

\section{LOCV calculations for magnetized nuclear matter}
\label{sec:1}

Let us consider a pure homogeneous symmetric nuclear
matter in a time-independent magnetic field $B$ along the $z$ axis at zero temperature.
Our system is composed of the spin-up $(+)$ and spin-down $(-)$ nucleons.
We denote the number densities of spin-up and spin-down nucleons by
$\rho^{(+)}$ and $\rho^{(-)}$, respectively. The spin
polarization parameter, $\delta$, is also introduced by
\begin{eqnarray}
     \delta=\frac{\rho^{(+)}-\rho^{(-)}}{\rho},
 \end{eqnarray}
where $-1\leq\delta\leq1$, and $\rho=\rho^{(+)}+\rho^{(-)}$ is the
total density of  system.

In order to calculate the energy of this system, we use LOCV method
as follows. We consider a trial many-body wave function of the form
\begin{eqnarray}
     \psi=F\phi,
 \end{eqnarray}
where $\phi$   is the uncorrelated ground-state wave function of $A$
independent nucleons and $F$ is a proper $A$-body correlation
function. Using Jastrow ansatz \cite{Jastrow}, $F$ can be
replaced by
\begin{eqnarray}
    F=S\prod _{i>j}f(ij),
 \end{eqnarray}
 where $S$ is a symmetrizing operator. Now, we consider a cluster expansion of the
 energy functional up to the two-body term,
 \begin{eqnarray}\label{tener}
           E([f])=\frac{1}{A}\frac{\langle\psi|H|\psi\rangle}
           {\langle\psi|\psi\rangle}=E^p _{1}+E^n _{1}+E _{2},
 \end{eqnarray}
 where $E^p _{1}$ and $E^n _{1}$ are the one-body energies of protons and neutrons, respectively, and $E _{2}$
 is the two-body energy.
At zero temperature, the one-body term for the protons, $E^p _{1}$, is
\begin{eqnarray}\label{E1p}
 E^p _{1}&=&\frac{e B}{\pi h c \rho}\sum_{i=+,-}\sum_{j^{(i)}=0}^{j^{(i)}_{max}}
 \{ \frac{1}{6}\frac{\hbar^{2}{k_F^{p(i)}}^2}{2m}+\frac{e \hbar B}{2m
         c}(j^{(i)}+\frac{1}{2})\nonumber \\&&-\lambda_i \mu_p B \}k_F^{p(i)}.
 \end{eqnarray}
Here, $j^{(i)}=0,\ 1,\ 2,\ 3,\ ...$ are the integers labeling the Landau levels \cite{Pathria} for a proton with spin projection $i$, $j^{(i)}_{max}$ is the largest integer, $k_F^{p(i)}$ is the proton Fermi momentum,
$\lambda_\pm=\pm1$, and $\mu_p$ is the proton magnetic moment. Besides, $e$ is the proton charge and $c$
is the speed of light.  $k_F^{p(i)}$ and $j^{(i)}_{max}$
are related to the proton density by
\begin{eqnarray}\label{rop}
 \rho_p=\frac{e B}{\pi h c} \sum_{i=+,-} \sum_{j^{(i)}=0}^{j^{(i)}_{max}} k_F^{p(i)}.
 \end{eqnarray}
The first two terms of Eq. (\ref{E1p}) correspond to the coupling between proton charge and magnetic field.
The last term of Eq. (\ref{E1p}) shows the interaction of magnetic field
with the proton magnetic moment.
The one-body term for the neutrons, $E^n_{1}$, can be written as
 \begin{eqnarray}
 \label{E1n}
E_{1}^n=\frac{1}{6\pi^2 \rho} \sum_{i=+,-} (\frac{3}{5}\frac{\hbar^{2}{k_F^{n(i)}}^2}{2m}-\lambda_i \mu_n B){k_F^{n(i)}}^3,
\end{eqnarray}
where $\mu_n$ is the neutron magnetic moment. In the above equation, $k_F^{n(i)}$ denotes the neutron Fermi momentum, and it
is related to the neutron density by
\begin{eqnarray}
 \rho_n=\frac{1}{6\pi^2} \sum_{i=+,-} {k_F^{n(i)}}^3.
 \end{eqnarray}
The first term of Eq. (\ref{E1n}) refers to the kinetic energy of neutrons and the second term
shows the interaction of the magnetic field
with the neutron magnetic moment.
It should be noted that for symmetric nuclear matter, we have $\rho_p=\rho_n=\rho/2$.

The two-body energy, $E _{2}$, is
\begin{eqnarray}
    E_{2}&=&\frac{1}{2A}\sum_{ij} \langle ij\left| \nu(12)\right|
    ij-ji\rangle,
 \end{eqnarray}
where
$$\nu(12)=-\frac{\hbar^{2}}{2m}[f(12),[\nabla
_{12}^{2},f(12)]]+f(12)V(12)f(12).$$
In the above equation, $f(12)$
and $V(12)$ are the two-body correlation function and inter-nucleon
potential, respectively. In LOCV formalism, the two-body correlation function, $f(12)$, induced by the strong nuclear force, is given by $f(12)=\sum^3_{k=1}f^{(k)}(r_{12})P^{(k)}_{12},$ where $P^{(k)}_{12}$ for $k=1-3$ are given in
Refs. \cite{Bordbar83,Bordbar718}.
Using this two-body correlation function and
the $AV_{18}$ two-body potential \cite{Wiringa}, after doing some algebra, we get an equation
for the two-body energy with similar form as given in
Ref. \cite{Bordbar76}. We note that this our new equation for two-body energy depends on the magnetic field in
an indirect way unlike our previous field free two-body energy.
In the next step, we minimize the two-body energy with respect to the
variations in functions $f^{(i)}$ subject to the
normalization constraint, $\frac{1}{A}\sum_{ij}\langle ij| h_{S_{z}}^{2}
-f^{2}(12)| ij\rangle _{a}=0$ \cite{Bordbar57}.
The function
$h_{S_{z}}(r)$ is defined as $$h_{S_{z}}(r)=[1-\frac{9}{2}(
\frac{J_{J}^{2}
(k_{F}^{(i)}r)}{k_{F}^{(i)}r}) ^{2}] ^{-1/2}$$ and $h_{S_{z}}(r)=1$ for $S_{z}=\pm1$ and $S_{z}= 0$, respectively. %
From the minimization of two-body cluster energy, we get a set of
coupled and uncoupled differential equations which have the same form as
those presented in Ref. \cite{Bordbar57}. By
solving these differential equations, we can obtain the correlation
functions to compute the two-body energy term.

\section{Results and discussion }\label{NLmatchingFFtex}
\label{sec:2}

In order to study the saturation properties of nuclear matter, we calculate the
energy up to the two-body term. Accordingly, we use Eq. (\ref{rop}) to compute the maximum Landau level.
Our results for the maximum Landau level, $j_{max}^{(i)}$, for the protons at different magnetic fields and densities have been given in Table~\ref{table1}. At lower magnetic fields, a large number of Landau levels are populated, and the nuclear matter acts as a system with zero magnetic field. However, in the strong magnetic fields, protons occupy only a few Landau levels.
While, in high density nuclear matter, protons can occupy a large number of Landau levels. Indeed, at high densities, even if the magnetic field is very strong, protons will occupy several Landau levels. Therefore, the effect of Landau quantization is more significant at lower densities. The results in Table~\ref{table1} imply that the main effect of Landau quantization is at lower densities when only one Landau level is occupied. It should be noted that in our calculations, the maximum Landau levels for the spin up and spin down protons are similar.

In figure 1, we have presented our saturation density
of nuclear matter as a function of the magnetic field
{with and without anomalous magnetic moments (AMM).}
In our previous study \cite{Bordbar57}, we obtained a value of $\rho_0=0.310\ fm^{-3}$
for the saturation density at zero magnetic fields. By increasing the magnetic field up to  $B\simeq 3 \times 10 ^ {17}\ G$,
{the saturation density gradually decreases and reaches the value $\rho_0=0.309\ fm^{-3}$ in both cases with and without AMM.} However, for $B>  3 \times 10 ^ {17}\ G$, the saturation density increases by increasing
the magnetic field. This incremental change is also reported
in Refs. \cite{Chakra7,Diener}. {Our results indicate that for high
magnetic fields, the saturation density is smaller for
the case including the nucleon AMM.}

{The magnetic field dependence of saturation energy of nuclear matter with and without anomalous magnetic moments (AMM) has been shown in figure~\ref{fig:7j}.}
Our results for the saturation energy at zero magnetic field has been shown to be $E_b=-18.46\ MeV$ \cite{Bordbar57}.
We have found that for the magnetic fields $B<  3 \times 10 ^ {17}\ G$, the saturation
energy slowly decreases by increasing the magnetic field, {and it approaches $E_b=-18.48\ MeV$ and $E_b=-18.47\ MeV$ in the cases with and without AMM, respectively.} However, for $B>  3 \times 10 ^ {17}\ G$, the saturation energy grows as the magnetic field increases. This behavior has been also reported in Ref. \cite{Diener}. It is obvious that the nuclear matter with AMM is more bound comparing to the case without AMM. Figure~\ref{fig:7j} also demonstrates a large increase in the saturation energy at $B\simeq 8 \times 10 ^ {18}\ G$. The growth of saturation energy with magnetic field shows that the nuclear matter becomes less bound when the magnetic field is present. It should be noted that although for $B>  3 \times 10 ^ {17}\ G$, the magnetic field leads to less binding in nuclear matter, it affects the nuclear matter so it becomes saturated at higher densities.
{Our results for the saturation energy at $B=10 ^ {19}\ G$, i.e. $E_b=-17.63\ MeV$ and $E_b=-16.63\ MeV$ with and without AMM, implies that even at the strongest magnetic field, the nuclear matter remains bound.}
It was concluded in Ref. \cite{Diener} that at $B \simeq 3 \times 10 ^ {17}\ G$, the system becomes unbound.
Conversely, the results obtained in Ref. \cite{Chakra7} confirm that for all magnetic fields studied, the saturation energy decreases by increasing the magnetic field, and the system turns out to be more bound.

{In figure~\ref{fig:8j1}, we have plotted our results for the spin and orbital magnetization of nuclear matter  ($M_{spin}$ and $M_{orbital}$, respectively) at the saturation density versus the
magnetic field.} {We have found that for $B<  3 \times 10 ^ {17}\ G$, the magnitude of the saturation magnetization
is nearly zero.} However, for $B>  3 \times 10 ^ {17}\ G$, this
quantity grows with the growth of the magnetic field. We can see that by increasing the magnetic field, the magnitude of orbital magnetization reaches the higher values comparing to the spin magnetization.
{In figure~\ref{fig:8j2}, the total magnetization of nuclear matter for two cases including AMM, i.e. $M_t=M_{spin}+M_{orbital}$,  and without AMM, i.e. $M_t=M_{orbital}$, are presented. It is clear that in the case without AMM, the nuclear matter takes the higher values for the magnitude of the magnetization.}

{The contribution of nucleon anomalous magnetic moments (AMM) in the magnetic susceptibility at the saturation density, i. e.}
\begin{eqnarray}
   \chi_{spin}(\rho,B)=
(\frac{\partial M_{spin}(\rho,B)}{\partial B})
_{\rho_0},
 \end{eqnarray}
{is shown in figure~\ref{fig:101j}.} {The contribution of AMM in the magnetic susceptibility
decreases with the magnetic field. In addition,  figure~\ref{fig:102j} shows the total magnetic susceptibility of nuclear matter at the saturation density, i. e.}
\begin{eqnarray}
   \chi(\rho,B)=
(\frac{\partial M_{t}(\rho,B)}{\partial B})
_{\rho_0}.
 \end{eqnarray}
{It should be noted that because of the small value of AMM contribution in the magnetic susceptibility (figure~\ref{fig:101j}), the magnetic susceptibility of nuclear matter with and without AMM are nearly similar. The total magnetic susceptibility of nuclear matter increases by increasing the magnetic field.} The weak negative magnetic susceptibility indicates
the diamagnetic properties of nuclear matter induced by the magnetic field \cite{Getzlaff}. Evidently, the effect of Landau diamagnetism on the magnetic susceptibility is more significant at low magnetic fields. For example, at  $B= 10 ^ {17}\ G$, we obtained $\chi/A=-0.343\times10^{-32}\ G^{-1}$. At higher magnetic fields, i.e. $B>  3 \times 10 ^ {17}\ G$, this quantity approaches  $\chi \simeq 0$ as the magnetic field increases.

The incompressibility of nuclear matter at the saturation density in the presence of a magnetic field which indicates the stiffness of equation of state (EOS) is given by
\begin{eqnarray}
\mathcal{K}(B)=9 \rho_0^2(B) [\frac{\partial^2E(\rho,B)}{\partial \rho^2}]_{\rho=\rho_0(B)}.
 \end{eqnarray}

{Figure~\ref{fig:9j} presents the incompressibility of nuclear matter
with  and without  anomalous magnetic moments as a function of the magnetic field.}
In Ref. \cite{Bordbar57}, we obtained
a value of $\mathcal{K}=302\ MeV$ for the incompressibility
{of symmetric nuclear matter at zero magnetic field.
Our results indicate that for low magnetic fields, in both cases with
and without AMM, the incompressibility decreases as the magnetic field increases.}
This behavior clearly shows the softening
of EOS at low magnetic fields. Besides, for magnetic fields about $B>  3 \times 10 ^ {17}\ G$,
the incompressibility is an increasing function of the magnetic
field, indicating the stiffening of EOS at high magnetic fields.
We can see that at each magnetic field, the incompressibility of nuclear matter
without AMM shows larger values comparing to the case with AMM.
It should be noted here that the softening and stiffening of
EOS are physical consequences of two competing factors,
the Landau quantization and the magnetization of system, respectively.
In fact, the increasing in the
degeneracy factor, $\frac{e B}{\pi h c}$, due to the increasing in the magnetic
field, leads to the softening of EOS.
{On the other hand, the magnetization of nuclear matter which originates from
the anomalous magnetic moments and the orbital motion of the nucleons (figure~\ref{fig:8j1})
leads to the stiffening of EOS. Our results confirm that at low magnetic
fields, with small magnetization (figure~\ref{fig:8j2}), the softening caused by Landau quantization
is the dominant effect. However, for the magnetic fields about $B>  3 \times 10 ^ {17}\ G$, in which the magnetization grows (figure~\ref{fig:8j2}), the softening of EOS is overwhelmed by stiffening due to the magnetization of nucleon matter,} as reported in Refs. \cite{Brod0,Yue}. {Similarly, the higher value of magnetization in the case without AMM (figure~\ref{fig:8j2})
leads to the larger values of incompressibility for nuclear matter without AMM.}
The incremental change in the incompressibility of nuclear matter has been also reported in Ref. \cite{Diener}. However,
in Refs. \cite{Chakra7,zhang}, the calculations result in the softening of EOS for all magnetic fields.

\section{Summary and Conclusions}
\label{sec:3}
In summary, we have investigated the effects of strong
magnetic fields on the cold symmetric nuclear matter satu
ration properties using LOCV method applying $AV_{18}$ two body
potential. {It was shown that for $B>  3 \times 10 ^ {17}\ G$, the
saturation density of symmetric nuclear matter with and without
nucleon anomalous magnetic moments increases
as the magnetic field grows.} In addition, we found that for
$B>  3 \times 10 ^ {17}\ G$, the saturation energy grows with growth
of magnetic field and consequently the nuclear matter becomes
less bound in the presence of magnetic field.
{It was confirmed that the orbital magnetization of the nuclear
matter has the main contribution in the magnetization of the nuclear matter.}
Our results for the magnetic susceptibility show that the nuclear
matter exhibits diamagnetic behavior particularly
at low magnetic fields. { Moreover, we observed the softening
of EOS caused by Landau quantization at low magnetic
fields, i.e. $B<  3 \times 10 ^ {17}\ G$, but we indicated that for
$B>  3 \times 10 ^ {17}\ G$, the softening is overwhelmed by stiffening
due to the magnetization of the nuclear matter.}

\acknowledgements{
This work has been supported financially by the Center for Excellence in Astronomy and Astrophysics (CEAA-RIAAM).
We also wish to thank the Shiraz University Research Council.}

\begin{table*}
\caption{Maximum Landau level for the protons at different magnetic
fields, $B$, for different values of density, $\rho$.}
\label{table1}       
\begin{center}  {\footnotesize
\begin{tabular}{|c|c|c|c|c|}
\hline& \multicolumn{4}{|c|}{Maximum Landau level}
\\\hline $\rho\ (fm^{-3})$  & \multicolumn{1}{c|}{$B=10^{17}\ G$} &
\multicolumn{1}{c|}{$B=10^{18}\
G$}&
\multicolumn{1}{c|}{$B=5\times10^{18}\ G$} &
\multicolumn{1}{c|}{$B=10^{19}\ G$}\\\hline
$0.05$ &26 &2 &0 &0    \\
$0.10$ &42 &3  &0& 0    \\
$0.15$ &55&5 &0 &  0   \\
$0.20$ &67 &6 &0 & 0  \\
$0.25$ &78&7 &1 & 0  \\
$0.30$ &88&8 &1 & 0   \\
$0.35$ &97&9 &1 &0  \\
$0.40$ &107&10 &1 &  0   \\
$0.45$ &115 &11  &1 & 0   \\
$0.50$ &124&12 &2&  1 \\
\hline
\end{tabular} }
\end{center}
\vspace*{2cm}  
\end{table*}

\newpage
\begin{figure*}
\vspace*{5cm}       
\includegraphics{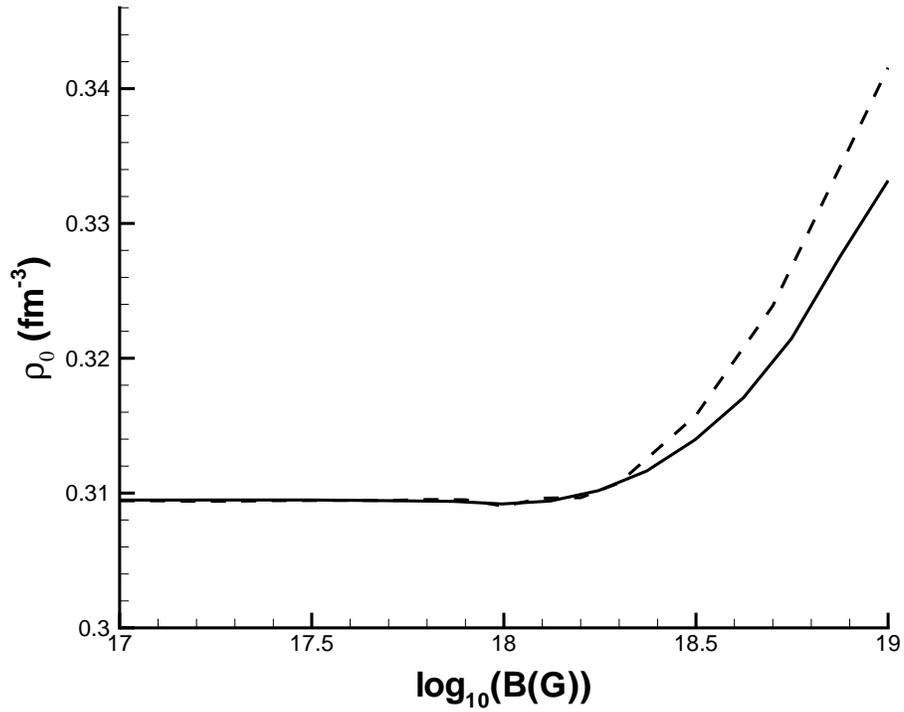}
\caption{{Saturation density of symmetric nuclear matter with (solid line) and without (dashed line) ANM versus the magnetic field.}}
\label{fig:6j}       
\end{figure*}
\newpage
\begin{figure*}
\vspace*{5cm}       
\includegraphics{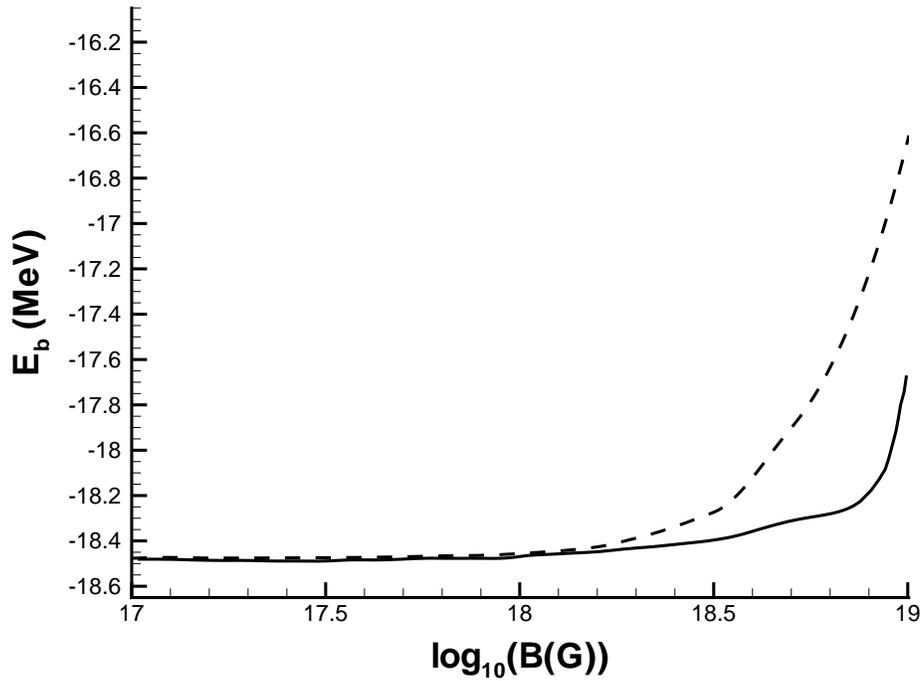}
\caption{As Fig. \ref{fig:6j} but for the saturation energy.}
\label{fig:7j}       
\end{figure*}

\newpage
\begin{figure*}
\vspace*{5cm}       
\includegraphics{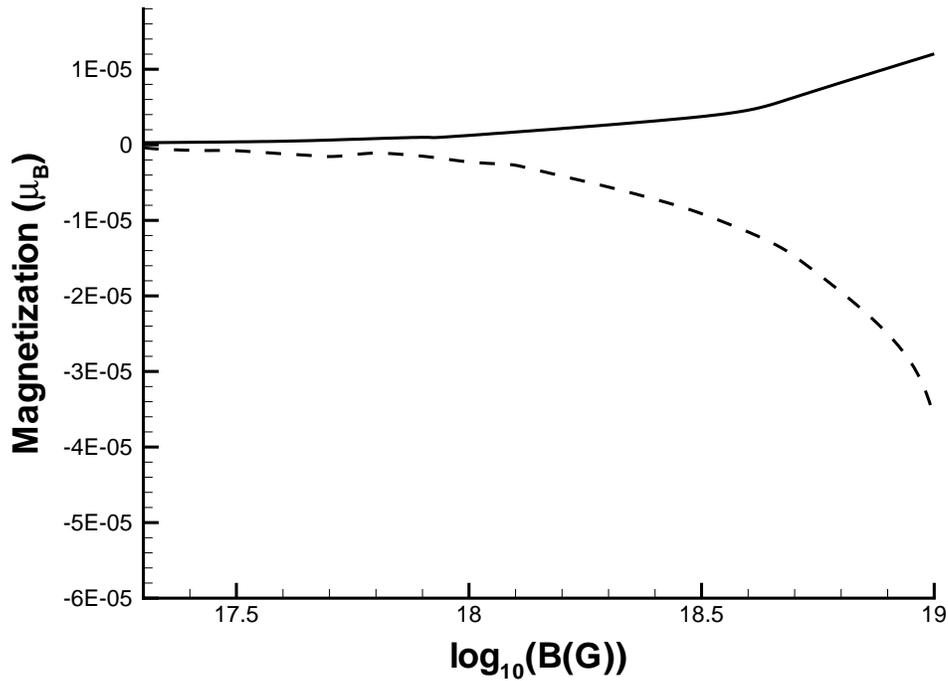}
\caption{{Spin magnetization (solid line) and orbital magnetization (dashed line)  of nuclear matter
at the saturation density versus the
magnetic field.}}
\label{fig:8j1}       
\end{figure*}

\newpage
\begin{figure*}
\vspace*{5cm}       
\includegraphics{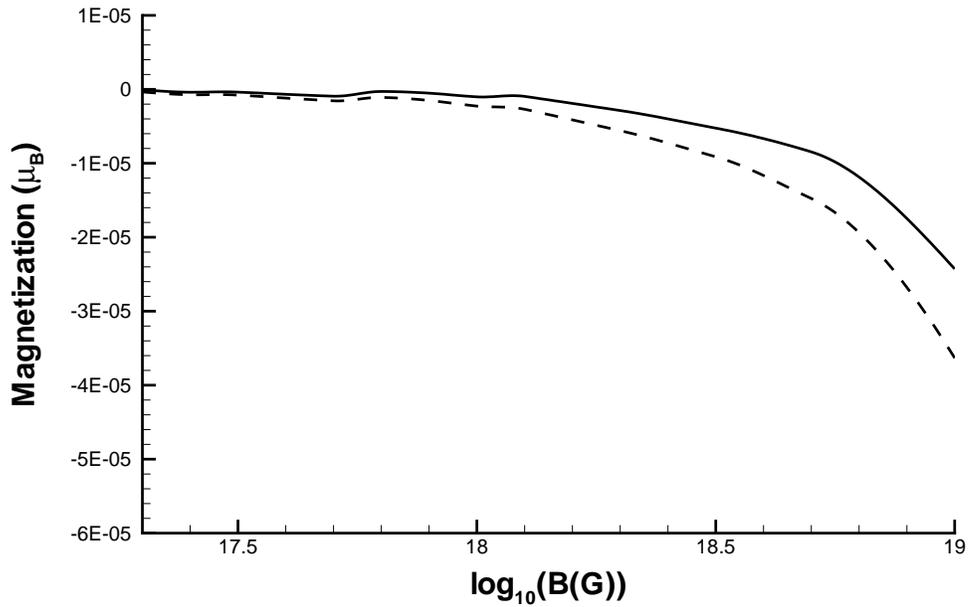}
\caption{{Total magnetization of nuclear matter with (solid line) and without (dashed line) ANM
 as a function of the magnetic field.}}
\label{fig:8j2}       
\end{figure*}

\newpage
\begin{figure*}
\vspace*{5cm}       
\includegraphics{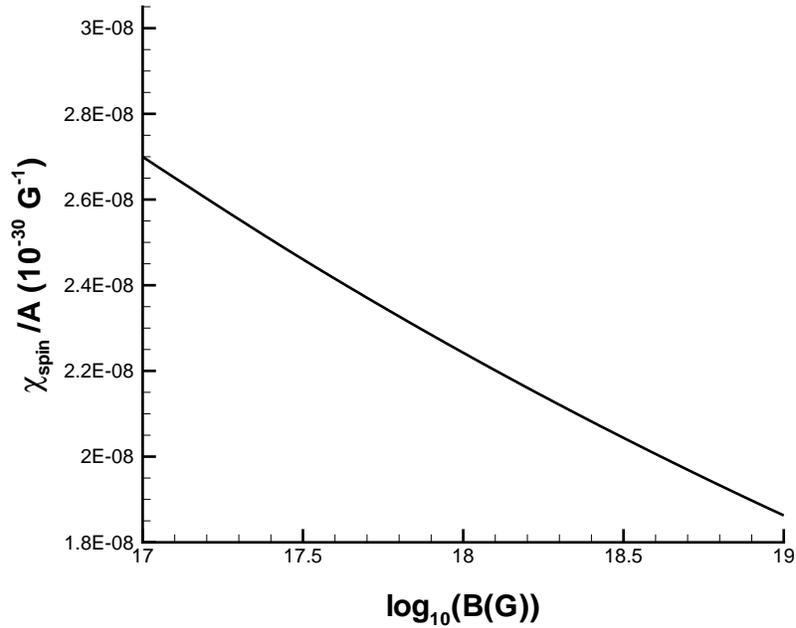}
\caption{{AMM contribution in the magnetic susceptibility of nuclear matter versus the
magnetic field. }}
\label{fig:101j}       
\end{figure*}

\newpage
\begin{figure*}
\vspace*{5cm}       
\includegraphics{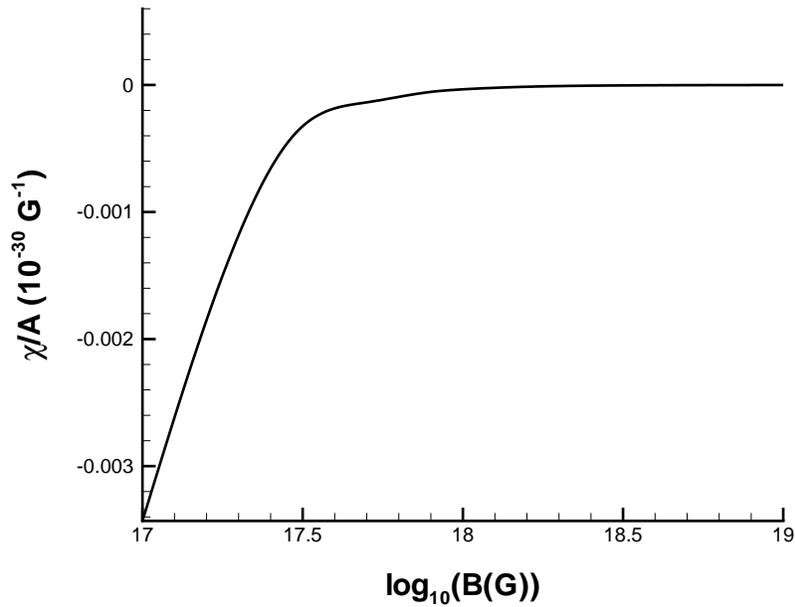}
\caption{{Magnetic susceptibility  of nuclear matter versus the
magnetic field. Due to the small value of AMM contribution in the magnetic susceptibility,
the magnetic susceptibility of nuclear matter with and without AMM are nearly identical.}}
\label{fig:102j}       
\end{figure*}
\begin{figure*}
\vspace*{5cm}       
\includegraphics{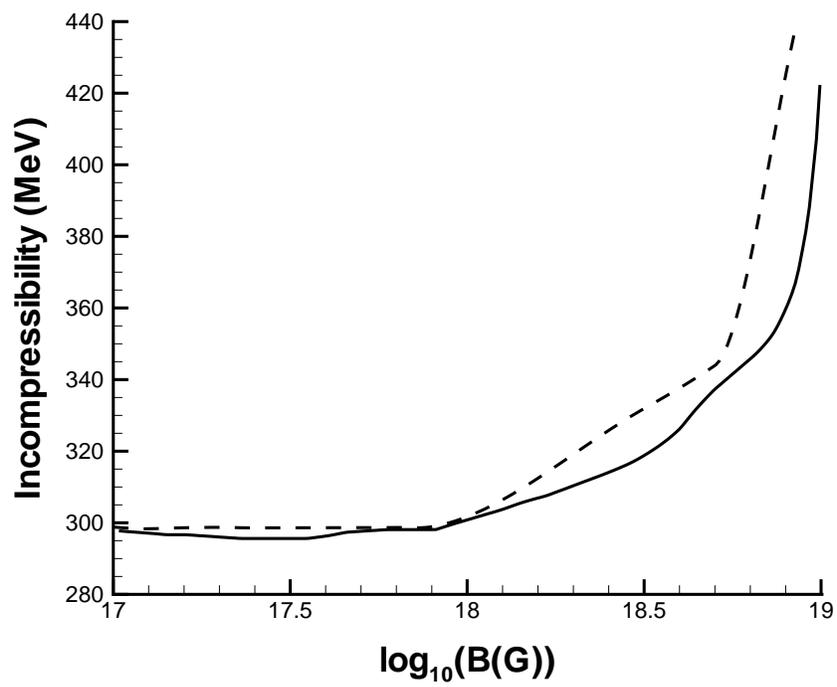}
\caption{As Fig. \ref{fig:6j} but for the incompressibility. }
\label{fig:9j}       
\end{figure*}



%

\end{document}